\documentclass[10pt,journal,twocolumn,twoside]{IEEEtran} 
\IEEEoverridecommandlockouts
\usepackage{setspace}
\usepackage{subfigure}
\usepackage{graphicx}
\usepackage{epstopdf}
\usepackage{float}
\usepackage{amsmath}
\usepackage[ruled]{algorithm2e}
\usepackage{algpseudocode}
\usepackage{array}
\usepackage{amsthm}
\usepackage{amsmath}
\usepackage{amssymb}
\usepackage{mdwmath}
\usepackage{mdwtab}
\usepackage{eqparbox}
\usepackage{stfloats}
\usepackage{hyperref}
\usepackage{cleveref}
\usepackage{fixltx2e}
\usepackage{cases} 
\usepackage{caption}
\usepackage{graphicx}
\usepackage{float} 

\usepackage{flushend}

\usepackage{xcolor}
\usepackage{makecell}

\usepackage{url}
\usepackage{cite}
\usepackage{booktabs}
\usepackage{multirow}
\allowdisplaybreaks[4]

\usepackage{wrapfig}

\makeatletter
\renewcommand{\maketag@@@}[1]{\hbox{\m@th\normalsize\normalfont#1}}%
\makeatother

\begin{document}

\title{\huge No Vision, No Wearables: 5G-based 2D Human Pose Recognition with Integrated Sensing and Communications\\
}

\author{Haojin Li, Dongzhe Li, Anbang Zhang, Wenqi Zhang\\
Chen Sun, \emph{Senior Member, IEEE} and Haijun Zhang, \emph{Fellow, IEEE}
\thanks{\emph{Corresponding author: Chen Sun}} %
\thanks{Haojin Li, Dongzhe Li and Haijun Zhang are with University of Science and Technology Beijing, China, Beijing 100083, China (E-mails: haojin.li@sony.com, m202420975@xs.ustb.edu.cn, haijunzhang@ieee.org, ).\\
Anbang Zhang is with Shandong University (E-mails: zab\_0613@163.com). \\
H. Li, Wenqi Zhang and C. Sun are with the Sony China Research Laboratory, Beijing 100027, China (E-mails: Haojin.Li@sony.com, Wenqi.Zhang@sony.com, Chen.Sun@sony.com).} %
\vspace{-10mm}
}
\maketitle

\begin{abstract}
With the increasing maturity of contactless human pose recognition (HPR) technology, indoor interactive applications have raised higher demands for natural, controller-free interaction methods. However, current mainstream HPR solutions relying on vision or radio-frequency (RF) (including WiFi, radar) still face various challenges in practical deployment, such as privacy concerns, susceptibility to occlusion, dedicated equipment and functions, and limited sensing resolution and range. 5G-based integrated sensing and communication (ISAC) technology, by merging communication and sensing functions, offers a new approach to address these challenges in contactless HPR. We propose a practical 5G-based ISAC system capable of inferring 2D HPR from uplink sounding reference signals (SRS). Specifically, rich features are extracted from multiple domains and employ an encoder to achieve unified alignment and representation in a latent space. Subsequently, low-dimensional features are fused to output the human pose state. Experimental results demonstrate that in typical indoor environments, our proposed 5G-based ISAC HPR system significantly outperforms current mainstream baseline solutions in HPR performance, providing a solid technical foundation for universal human-computer interaction.
\end{abstract}


\IEEEpeerreviewmaketitle

\section{Introduction}\label{S1}

Integrated sensing and communication (ISAC) stands as a core enabling technology for the 5G-advanced and 6G eras \cite{11145172}. By seamlessly converging wireless communication and radar sensing functionalities, it realizes the visionary principle of multiple uses per frequency band and dual capabilities per network without occupying additional spectrum resources and by reusing existing base station (BS) hardware infrastructure. This breakthrough enables wireless systems to deliver high-speed, low-latency data transmission while simultaneously performing precise sensing, such as target localization, pose recognition, without introduced additional spectrum or dedicated hardware infrastructure. Among its promising applications, human pose recognition (HPR) stands out as a critical enabler for smart homes, assisted living, and entertainment. Traditional methods, however, are often hampered by environmental constraints, privacy issues, or hardware costs. In recent years, alternative sensing pathways including vision-based, WiFi-based, and radar-based techniques have diversified the technological landscape. We will conduct a systematical review to examine the state of the art across these domains and spotlights the emerging potential of 5G-based HPR.

Vision-based HPR leverages red, green, blue (RGB), depth (RGBD), binocular, or omnidirectional cameras. Early HPR work depended on handcrafted features and classifiers \cite{1467360}, the field was revolutionized by the advent of convolutional neural networks (CNNs). The first CNN-based model broke away from manual feature design, unlocking higher accuracy and scalability to multi-person scenarios\cite{6909610}. Later, methods like part affinity fields (PAFs) enabled real-time multi-person 2D pose estimation by solving keypoint association challenges\cite{8765346}. Recently, transformer-based architectures have embedded keypoints as learnable tokens, capturing appearance cues and structural constraints while reducing model complexity\cite{9710987}. To further enhance robustness in dynamic environments, hierarchical contrastive consistency constraints have been introduced to model relationships between keypoints and body parts across multiple feature levels\cite{11138023}.

As a prominent category of RF-based sensing, WiFi-based HPR systems function by exploiting the perturbations in channel state information (CSI) that result from human activities, thereby offering inherent advantages in occlusion penetration and privacy preservation. Early systems often overlooked phase information, limiting their capability. Breakthroughs now allow dense pose estimation from WiFi signals even under occlusion and in multi-person settings\cite{geng2022denseposewifi}. Recent transformer-based frameworks analyze long-sequence CSI data, and curated datasets such as Wi-Pose have accelerated progress\cite{e25010020}. By deploying multiple WiFi devices and adopting end to end transformer architectures, researchers have achieved notable improvements in spatial resolution and multi-person tracking\cite{10656837}.

Radar-based HPR systems forms the core of advanced RF-based sensing, providing the fine-grained motion capture. Some approaches employ cross-modal supervision, where visual poses guide wireless signal learning\cite{8578866}. Others transform RF signals directly into pose representations via optimal transport theory, operating at the feature level rather than the signal level\cite{63d9d87490e50fcafd57e656}. To reduce computational load, antenna arrays can first localize subjects on RF heatmaps before performing refined pose estimation on cropped regions\cite{10250968}. For sparse radar point clouds, mapping distance, elevation, azimuth, and signal strength into RGB-like channels enables effective CNN-based 3D skeleton prediction\cite{9083948}.

While vision-based systems offer high resolution, they struggle with lighting changes and occlusion. WiFi-based systems are low-cost and easy to deploy but are bandwidth limited, affecting accuracy in complex poses. Radar-based systems penetrate obstacles and work in the dark, yet often suffer from limited range and spectrum inefficiency.

5G-based HPR emerges as a transformative solution, natively built upon the ISAC architecture. This paradigm enables the deep reuse of hardware and spectrum resources, striking a balance among non-intrusiveness, interference robustness, and wide-area coverage while meeting stringent demands for high precision, resolution and low latency. The high frequency band ensures sensing resolution. Its passive, wear-free design boosts user acceptance. Critically, by analyzing poses through wireless signals rather than visual data, it significantly mitigates privacy risks. Moreover, it leverages existing 5G infrastructure and authorized spectrum, avoiding extra hardware or dedicated frequency resources. Despite these strengths, research into 5G-based HPR remains scarce.

In this paper, a systematic review of HPR technologies is presented, with a dedicated focus on the emerging 5G-based paradigm. First, four mainstream approaches, Vision, WiFi, radar, and 5G-based are analyzed and compared. Subsequently, the key use cases of 5G-based HPR in indoor environments are surveyed. Building on these insights, a novel system architecture for 5G-based HPR is proposed. To validate its feasibility, an experimental prototype was implemented and evaluated against established benchmarks, demonstrating its competitive performance and practical potential.

\begin{figure*}[t]
\centering
\includegraphics[width=0.95\linewidth]{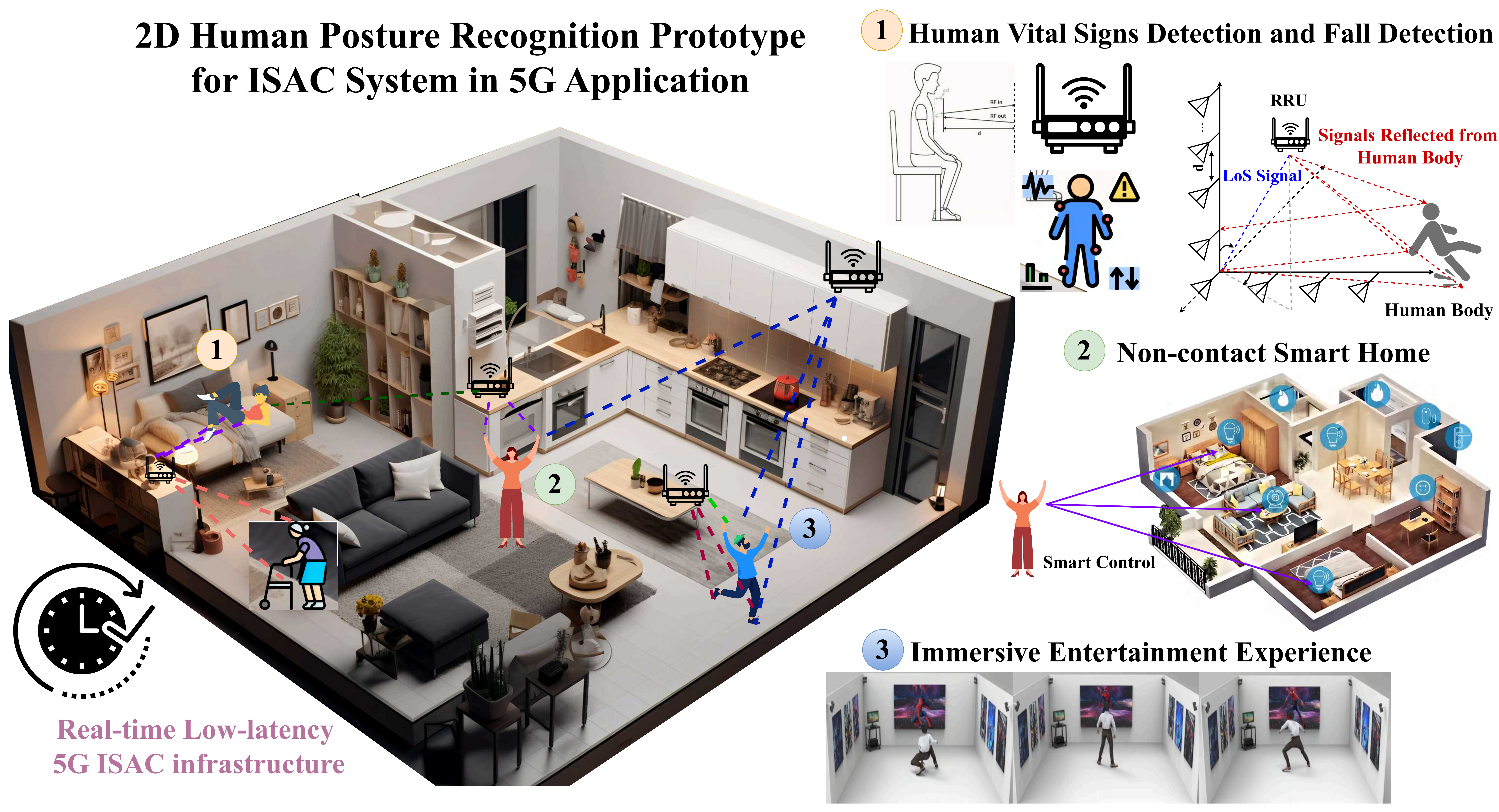}
    \caption{5G-based 2D human pose recognition prototype supported with real-time, low-latency.  {\small{\textcircled{\scriptsize{1}}}} Human vital-sign sensing and fall detection: RF echoes from the human body are used to monitor respiration, pose stability, and accidental falls. {\small{\textcircled{\scriptsize{2}}}} Non-contact smart home control: Recognized poses and gestures drive device control and scene automation across multiple rooms. {\small{\textcircled{\scriptsize{3}}}} Immersive entertainment interaction: Full-body tracking enables controller-free gaming and interactive multimedia experiences.}
\label{fig1} 
\end{figure*}

\section{Application Use Cases}

In this section, we explore real-world indoor applications of the 5G-based HPR system. Using the controller-free 2D prototype shown in Fig. \ref{fig1} as a reference, we detail three key scenarios, which include immersive entertainment, fitness guidance, and seamless smart home control.

\subsection{Entertainment Interaction: Immersive Experience }

Indoor digital entertainment increasingly demands more intuitive, and embodied interaction modalities. The 5G–based HPR system enables realistic, controller-free entertainment experiences by directly mapping users’ natural full-body movements to interactive commands. Motion-sensing games can therefore be operated through poses such as waving hands to trigger attacks, jumping to evade obstacles, or tilting the torso to change direction, without relying on handheld controllers or wearable sensors. This interaction paradigm enhances user enjoyment and inclusiveness, making it suitable for family entertainment, parent-child gaming, etc.

Beyond traditional gaming, the system further supports emerging virtual entertainment ecosystems. In live streaming and virtual reality (VR) applications, recognized pose can be synchronized with animated avatars or virtual idols, significantly improving viewers’ sense of immersion and participation. In indoor karaoke scenarios, body swings, rhythmic arm gestures, or dancing movements can dynamically control lighting, audio effects, or projection visuals, thereby enriching the atmosphere and promoting audience engagement.

\subsection{Fitness and Wellness: Scientific Guidance and Tracking}

Health-oriented applications require continuous, fine-grained monitoring of human movements to assess exercise performance, detect abnormalities, and improve training outcomes.  The proposed system provides real-time pose evaluation during home-based fitness practices such as yoga or body-weight workouts.  It analyzes joint angles, motion ranges, and movement stability to determine whether each pose is executed correctly, offering timely corrections through auditory or visual feedback.  This capability effectively transforms the system into a virtual personal trainer, enabling personalized exercise guidance without professional supervision.

For rehabilitation medicine, pose trajectories and joint motion characteristics serve as important clinical indicators. The system can monitor patients recovering from orthopedic injuries, neurological disorders, or post-surgical procedures by quantifying motion completion ratios, asymmetry, gait patterns, and balance control.  The collected data can be summarized into structured rehabilitation reports for clinicians, supporting remote assessment, treatment plan adjustments, and long-term progress tracking. Continuous pose analysis can detect the risk of falls, muscle weakness, or abnormal gait deterioration at an early stage, all of which are key predictors of functional decline. The system can automatically issue alerts to caregivers or medical institutions, improving safety for seniors living independently.

\subsection{Smart Home: Seamless Interaction and Automation}

As smart homes evolve toward cognitive and anticipatory environments, natural and frictionless interaction becomes essential. The 5G-based HPR system replaces conventional remote control or voice-based commands with intuitive pose driven controls. Users may wave to switch lights on or off, use a mute gesture to silence a television, or adjust indoor temperature with simple upward or downward arm motions. These interactions are particularly useful when hands are occupied, for example, during cooking, cleaning, or caregiving, thereby improving accessibility and convenience.

Beyond device-level operations, pose triggered scene automation enhances contextual awareness and living comfort. For instance, lying on a sofa may automatically activate a movie-watching mode, closing curtains, dimming lights, and turning on audiovisual equipment, while standing up may restore a daily mode for routine activities. This approach benefits elderly individuals and young children, who may struggle to remember or operate complex smart home interfaces.

\section{MiDiPose: 5G-based multi-feature domains for human pose recognition}
In this section, we propose a 5G-based multi-feature domains framework for HPR (MiDiPose) in Fig. \ref{fig2}.
The CSI and ground truth visual data are obtained in real measurements. 
MiDiPose extracts amplitude, phase, and Doppler information as complementary feature domains from CSI.
Then, the features are encoded and aligned into feature embeddings, which are fused to enable accurate HPR.
\subsection{Data Acquisition}
A distributed multi-node collaborative sensing system based on a 5G architecture is constructed. The transmitter of the system employs a commercial smartphone as the user equipment (UE). The receiver utilizes star-connected cell merging technology, where multiple remote radio units (RRUs) are merged into one logical cell via the baseband unit (BBU). Consequently, the UE and multiple RRUs collectively form an one transmitter and multiple receiver sensing model.  When human activity occurs within the sensing area, the wireless propagation channel changes accordingly. Channel estimation is then performed using the uplink SRS transmitted by the UE. At the BBU side, CSI from the links between different RRUs and the UE is obtained. The captured CSI thus corresponds to human pose.
\begin{figure}[!t]
    \centering
    \includegraphics[width=0.95\linewidth]{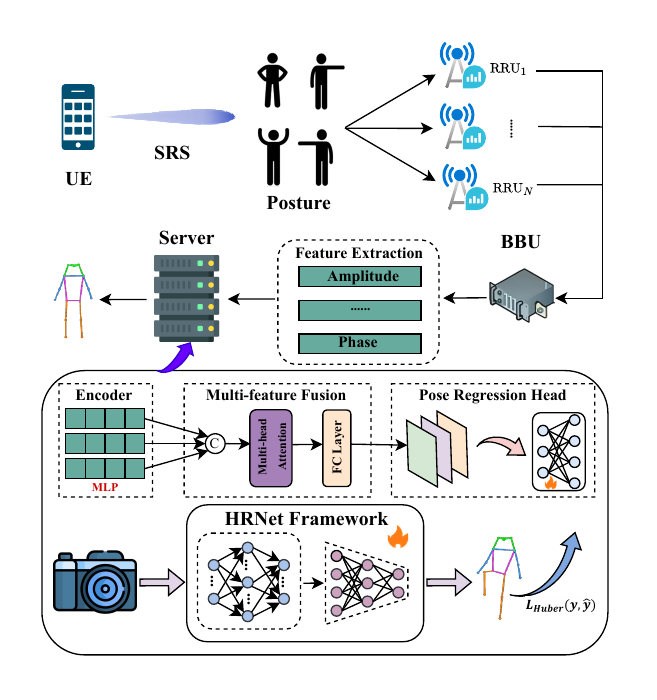}
    \caption{Schematic diagram of MiDiPose: 5G-based multi-feature domains for human pose recognition.}
    \label{fig2}
\end{figure}

Additionally, HRNet is an efficient image processing network that is widely used for HPR tasks. HRNet serves as the preprocessing module for the input images. Specifically, the input image is first passed through the HRNet model. The model then outputs a set of 2D keypoint coordinates that represent key body parts, such as the head, shoulders, elbows, knees, and so on. These keypoint coordinates are used as supervision labels to guide the MiDiPose model during the subsequent training process.

\subsection{Multi-feature Domains}

We aim to fully exploit the human motion information from CSI. Since human activities affect the energy, propagation distance, and frequency of wireless signals, these variations are correspondingly mapped onto the amplitude, phase, and Doppler frequency shift (DFS) components of CSI. By leveraging the rich physical properties of CSI, features are extracted from multiple domains, thereby providing a more comprehensive and information dense input representation for the subsequent neural network.

\subsubsection{Amplitude Domain}
After obtaining CSI, the system extracts the amplitude component from it.
Focusing on state recognition at a single timestamp, which lacks the overall features inherent in the temporal structure.
To address this limitation, the MiDiPose introduces a sliding time window mechanism, which computes the normalized standard deviation, median absolute deviation, and interquartile range of the amplitude data within the window. These statistical features are combined with the amplitude data at the current timestamp to provide richer contextual temporal information for the model.

\subsubsection{Phase Domain}
Regarding the processing of CSI phase information, the inherent range of the arctangent function, confined to $ ( -\pi/2, \pi/2 ) $, which causes the phase data to exhibit a wrapped morphology around the horizontal axis. This wrapping effect introduces abrupt discontinuities in the phase, which impedes the neural network's capacity to learn the latent feature information embedded within such discontinuities. Consequently, the phase information processing workflow is partitioned into two primary components, i.e., phase unwrapping and linear transformation. Furthermore, within the phase domain, phase differentials are derived from data acquired by disparate receiving RRUs, serving as a critical feature for multi-node collaborative sensing and thereby enhancing sensing performance.
\subsubsection{Doppler Domain}
When a person performs activities within the sensing region, the signal propagation paths between the UE and the RRUs undergo corresponding changes, which can be considered equivalent to relative motion between the transmitter and the receiver. This motion induces variations in the wavelength and frequency of the received signals, manifesting as DFS. DFS effectively characterizes the moving speed and direction of the target, serving as an essential feature for pose reconstruction. In this work, DFS is estimated based on the phase variations between adjacent timestamps.

\subsection{Multi-feature Fusion Model Architecture}
MiDiPose aims to map amplitude, phase, and Doppler domains into a shared semantic space for domain alignment and semantic consistency.
Then, it learns cross-domain interactions and outputs 2D human skeleton keypoints.
The architecture consists of three components.
\subsubsection{Feature Encoding and Alignment}
Different features have distinct statistics and physical meanings in amplitude, phase, and Doppler domains. Therefore, MiDiPose uses a light-weight multi-layer perceptron (MLP) encoder for each feature vector, as shown in \ref{fig2}. First, each encoder performs nonlinear compression and abstraction on features. It produces a low-dimensional latent vector that reduces scale mismatch and suppresses noise redundancy. Moreover, the encoded vectors from different domains are projected into the same latent representation space, which achieves semantic-level alignment.
As a result, the model obtains a more discriminative joint representation.
\subsubsection{Feature Fusion}
A basic fusion strategy is to concatenate the aligned feature embeddings from different domains along the feature dimension. This forms a joint feature for the downstream regression head. However, a simple concatenation cannot explicitly model complementarity and dependencies across domains. To improve cross-domain interaction modelling, MiDiPose further introduces a fusion module built with multi-head attention and an MLP. After alignment, features from different domains can be regarded as a set of feature tokens. The attention mechanism selectively focuses on features that are more sensitive to pose changes. The output of attention is then fed into an MLP, which allows nonlinear mixing and dimensional transformation.
\subsubsection{2D Human Pose Output}
MiDiPose feeds the fused feature representation into a pose regression head and generates 2D skeleton keypoint coordinates. Since the fused representation is low-dimensional, a lightweight convolutional or residual backbone can be adopted. For example, a ResNet backbone can apply deeper nonlinear transformations to strengthen the representation. Finally, an MLP regression head projects the enhanced latent features into the keypoint space and outputs the 2D coordinates of each joint.

\section{Experiments and Discussions}
\subsection{Experimental Settings}
\emph{\textbf{1) ISAC Prototype Platform:}}
The 5G ISAC experimental platform utilizes BS from H3C, with Sony Xperia 1 IV smartphones serving as UE. The core network employs the WX3540X, while the BS uses the BBU5200 Series. The transmitting smartphone and three receiving RRUs are each mounted on tripods and placed at four corners, forming a sensing area measuring $3.3~\text{m} \times 2.7~\text{m}$. All devices are positioned on the ground at a height of $1.5~\text{m}$ to better capture the movements and poses of the human body within the sensing area.

\emph{\textbf{2) Data Collection and Evaluation Protocol:}}
The system synchronously collects 5G CSI samples and RGB image frames. The parsed CSI is a tensor of size $\emph{n} \times 544 \times 3 \times 7$, where $\emph{n}$ represents the number of received CSI packets, 544 denotes the number of subcarriers, 3 corresponds to the number of receiving RRUs, and 7 indicates the number of extracted features. The sampling rates of the images and CSI are approximately $15~\text{Hz}$ and $25~\text{Hz}$, respectively. Subsequently, based on each image frame, the CSI data with the closest timestamp is selected to achieve precise data alignment.

A dataset targeting daily human activities is constructed, comprising a total of 221{,}410 synchronized frames. The dataset covers distinct activities, including stepping in place, squatting, hand raising, lunging, and walking back and forth.

The dataset was partitioned into training, testing, and validation sets with a ratio of 70\% : 20\% : 10\%. To quantify estimation accuracy, we adopt the percentage of correct keypoint (PCK) metric as defined in \cite{10152057}. Furthermore, our evaluation distinguishes between motion state and motion process: The former refers to specific static poses within an action, while the latter encompasses the continuous sequence of frames throughout the entire execution of the movement.

\emph{\textbf{3) Approaches for Evaluation:}}
We compare the proposed MiDiPose with state-of-the-art scheme MetaFi\cite{10152057}.

\begin{itemize}
\item \textbf{MiDiPose:} The model is implemented using PyTorch and trained for 100 epochs using the stochastic gradient descent (SGD) algorithm to optimize the loss function. We set the batch size to 64, the learning rate to 0.008, and the momentum to 0.9. The learning rate follows a multi-step decay strategy, where it is multiplied by a factor of 0.5 every 10 epochs.
\item \textbf{MetaFi:} A WiFi-based IoT-enabled human pose estimation scheme is proposed for metaverse avatar simulation. Specifically, a deep neural network is designed with customized convolutional layers and residual blocks to map the channel state information to human pose landmarks. We adapt the MetaFi method to 5G signals by using the same neural network architecture but with our 5G CSI data as input.
\end{itemize}

To facilitate fair comparisons, the dimensionality of the representations encoded by all methods is standardized. Furthermore, all models utilize the identical neural network backbone originally proposed in MetaFi, ensuring equivalent constraints on computational complexity and memory usage to simulate deployment on resource-limited edge devices.

\subsection{Experimental Results}

As shown in Tab. \ref{tab:performance_state}, marktime1 and marktime2 denote the instants when the left and right legs are lifted during in-place marching, respectively. The term motion state refers to the static pose at a specific moment of an action, whereas motion process in Tab. \ref{tab:performance_process} corresponds to the entire dynamic sequence of performing that action. Note that we employ the PCK metric to evaluate the all the scheme, reporting results under different strictness level, i.e., PCK@5, @10, @20, and @30. Here, PCK$@\alpha$ denotes a prediction as correct if its distance from the ground-truth is within $\alpha $\% of the torso length, with lower thresholds representing stricter accuracy criteria.

MiDiPose achieves consistently higher PCK on almost all motion states, especially under the strict PCK@5, as demonstrated in Tab. \ref{tab:performance_state}. For in-place marching, MiDiPose improves the PCK@5 on marktime1 and marktime2 from 64.71\% and 28.10\% (MetaFi) to 79.74\% and 93.46\%, respectively, gains of more than 15\% and 65\%. Similar advantages are observed for lunge1, lunge2 and squat, where MiDiPose yields $20\%\sim25\%$ improvements at PCK@5. As the threshold increases, MiDiPose quickly saturates to nearly perfect accuracy at PCK@20 and PCK@30, it reaches 100\% on all motion states, while MetaFi still suffers from noticeable errors on several states such as marktime2 and squat.

The motion processing in Table \ref{tab:performance_process} similarly exhibits this trend. At PCK@5, MiDiPose outperforms MetaFi by about 20\% on marktime, over 5\% on walk and squat, over 10\% on rise hand. This demonstrates its greater robustness on complete action sequences rather than isolated poses. 
At moderate thresholds (i.e., PCK@20 and PCK@30), MiDiPose maintains competitive or superior for all motions, achieving $99.73\%\sim100.00\%$ on Rise hand and $99.08\%\sim99.82\%$ on walk, while MetaFi is clearly behind on several motions, particularly marktime and squat. Overall, MiDiPose provides more accurate and stable pose estimation across both static and dynamic scenarios, which confirms the effectiveness of the proposed strategy compared with MetaFi.

\begin{table*}[htbp]
\centering
\caption{Performance comparison between MiDiPose and MetaFi under various PCK thresholds (Motion States).}
\label{tab:performance_state}
\resizebox{\textwidth}{!}{%
\begin{tabular}{llcccccccc}
\toprule
\multicolumn{2}{c}{\multirow{2}{*}{\textbf{Motion State (\%)}}} & Marktime1 & Marktime2 & Lunge1 & Lunge2 & Rise hand1 & Rise hand2 & Walk & Squat \\
\multicolumn{2}{c}{} & (\%) & (\%) & (\%) & (\%) & (\%) & (\%) & (\%) & (\%) \\
\midrule
\multirow{2}{*}{PCK5}  
& MiDiPose & \textbf{79.74} & \textbf{93.46} & \textbf{69.41} & \textbf{65.88} & 97.48 & 98.32 & 92.86 & 70.00 \\
& MetaFi & 64.71 & 28.10 & 41.18 & 42.35 & \textbf{98.29} & \textbf{98.76} & \textbf{95.17} & \textbf{71.69} \\
\midrule
\multirow{2}{*}{PCK10} 
& MiDiPose & \textbf{98.69} & \textbf{99.35} & \textbf{100.00} & \textbf{100.00} & \textbf{98.74} & \textbf{99.45} & 94.48 & 92.00 \\
& MetaFi & 87.58 & 50.98 & 91.76 & 94.12 & 98.69 & 99.21 & \textbf{95.62} & \textbf{93.01} \\
\midrule
\multirow{2}{*}{PCK20} 
& MiDiPose & \textbf{100.00} & \textbf{100.00} & \textbf{100.00} & \textbf{100.00} & \textbf{100.00} & \textbf{100.00} & \textbf{100.00} & \textbf{100.00} \\
& MetaFi & \textbf{100.00} & 65.36 & \textbf{100.00} & \textbf{100.00} & \textbf{100.00} & \textbf{100.00} & \textbf{100.00} & 97.06 \\
\midrule
\multirow{2}{*}{PCK30} 
& MiDiPose & \textbf{100.00} & \textbf{100.00} & \textbf{100.00} & \textbf{100.00} & \textbf{100.00} & \textbf{100.00} & \textbf{100.00} & \textbf{100.00} \\
& MetaFi & \textbf{100.00} & 73.86 & \textbf{100.00} & \textbf{100.00} & \textbf{100.00} & \textbf{100.00} & \textbf{100.00} & 98.16 \\
\bottomrule
\end{tabular}%
}
\end{table*}

\begin{table}[t]
\centering
\setlength{\tabcolsep}{4pt} 
\caption{Performance comparison between MiDiPose and MetaFi under various PCK thresholds (Motion Processes).}
\label{tab:performance_process}
\begin{tabular}{llcccccc}
\toprule
\textbf{Motion} & \textbf{Method} & PCK5 & PCK10 & PCK20 & PCK30\\
\midrule
\multirow{2}{*}{Marktime} 
    & MiDiPose & \textbf{63.09} & \textbf{84.41} & \textbf{97.21} & \textbf{99.85} \\
    & MetaFi   & 41.18          & 70.15          & 88.38          & 91.91          \\
\midrule
\multirow{2}{*}{lunge} 
    & MiDiPose & \textbf{50.14} & \textbf{76.19} & 88.52          & 91.60          \\
    & MetaFi   & 38.38          & 74.51          & \textbf{91.88} & \textbf{96.08} \\
\midrule
\multirow{2}{*}{Rise hand} 
    & MiDiPose & \textbf{68.72} & \textbf{90.64} & \textbf{99.73} & \textbf{100.00} \\
    & MetaFi   & 55.88          & 84.76          & 98.13          & 98.93           \\
\midrule
\multirow{2}{*}{Walk} 
    & MiDiPose & \textbf{75.92} & \textbf{95.22} & \textbf{99.08} & \textbf{99.82} \\
    & MetaFi   & 70.59          & 86.03          & 95.77          & 99.63          \\
\midrule
\multirow{2}{*}{Squat} 
    & MiDiPose & \textbf{90.34} & \textbf{98.74} & \textbf{100.00} & \textbf{100.00} \\
    & MetaFi   & 83.19          & 92.22          & 95.17           & 98.95          \\
\bottomrule
\end{tabular}
\end{table}

\begin{figure*}[!ht]\setcounter{figure}{2}
	\centering
	\includegraphics[width=0.95\linewidth]{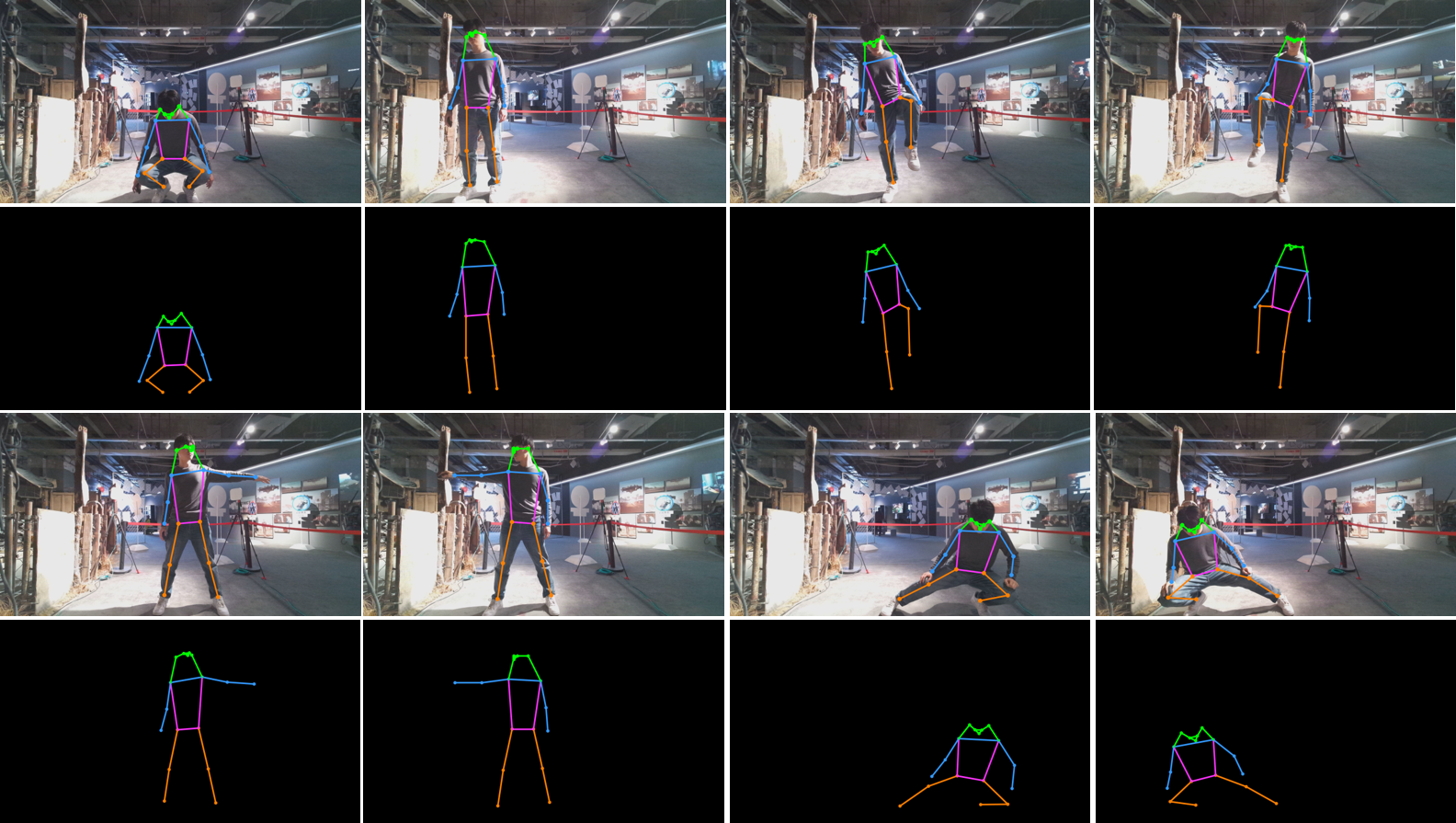}
	\caption{The human pose coordinates are generated by the visual model and our proposed MiDiPose model, respectively. The first and third rows present the results of the visual model, while the second and fourth rows illustrate those of the MiDiPose model.}
	\label{fig3} 
\end{figure*}

The superior accuracy of MiDiPose primarily stems from its multi-domain feature fusion architecture, which extracts complementary amplitude, phase, and Doppler features from CSI and aligns them within a shared semantic space. By explicitly modeling cross-domain interactions and dependencies through attention-based fusion, the model effectively captures both static postures and dynamic motion variations, thereby providing enhanced representational capability for fine-grained human actions.

Fig. \ref{fig3} provides a qualitative comparison between the human poses predicted by MiDiPose and a vision-based scheme. The first and third rows show the original RGB images overlaid with the skeletons produced by the visual model, while the second and fourth rows show corresponding skeletons estimated solely from MiDiPose. Each column corresponds to a different pose, including in-place squatting, upright standing, leaning, side stretching, and large range leg movements. Note that MiDiPose can accurately recover the overall body configuration across diverse poses. The positions of key joints, such as shoulders, hips, knees, and ankles, exhibit high alignment with visual model outputs.
Even in complex occlusion scenarios (e.g., deep squats, lateral bends, or extensive limb extensions), matching accuracy remains outstanding. CSI-based predictions of limb orientation and body contours also align closely with visual results, demonstrating that MiDiPose effectively captures fine-grained human pose features.

\section{Open Issues and Challenges}
\subsection{Diverse Human Poses}

The human structure is a complex articulatory system, characterized by high dimensionality with numerous freedom degrees. Its motion space is not a simple, predictable map. Instead, it exists as a fluid continuum punctuated by sudden, discrete shifts. The complexity of movements manifests across multiple levels, i.e., encompassing daily actions like walking, sitting, and waving, as well as specialized scenarios such as fitness and rehabilitation training, alongside unconventional postures like single-leg standing, bending to pick up objects, and body twisting. These postures often present as continuous temporal sequences rather than isolated static poses. Changes in acceleration and angular velocity during motion further amplify the complexity of the posture space. Moreover, traditional CNN and Transformer models heavily rely on large amounts of labeled data, lacking feature transfer capabilities for unseen postures, thereby resulting in insufficient model generalization.

\subsection{Cooperative Synchronization Among Multiple BSs}

Simultaneous reception of human echo signals by multiple BSs, coupled with joint estimation of position, orientation, and kinematic parameters, significantly reduces joint angle errors in pose estimation. However, synchronization accuracy remains the core constraint for multi-BS collaboration, i.e., ISAC HPR requires nanosecond-level time synchronization and 10~\text{Hz}-level frequency synchronization. The synchronization capabilities of existing communication BS cause ranging, angle measurement, and velocity measurement errors to accumulate with distance, severely compromising the consistency of attitude estimation. Related research can be pursued through three collaborative dimensions, i.e., adopting atomic-clock-grade crystal oscillators with dedicated synchronous RF links, designing synchronization frame structures tailored for ISAC, and developing synchronization error compensation models.

\subsection{Inherent Properties of Communication oriented Waveforms}
 
The core challenge of ISAC lies in the inherent trade-off between its dual essential requirements. possesses inherent advantages and limitations. Radar-centric waveforms, such as linear frequency modulation (LFM), frequency modulated continuous Wave (FMCW), and chirp, offer large time-bandwidth products and high sensing resolution, yet adopt simplistic modulation schemes. This results in weak data carrying capacity and low spectral efficiency. In contrast, communication-centric waveforms such as orthogonal frequency division multiplexing (OFDM), deliver high spectral efficiency, support multi-user parallel transmission, and achieve data rates up to the Gbps levels. However, they suffer from issues such as dispersed time domain signal energy, elevated sidelobe levels, inadequate sensing resolution, and performance degradation caused by the coupling between target parameter estimation and communication demodulation. Multi-waveform collaboration addresses these limitations by realizing decoupling, complementarity and optimization of communication and sensing performance through waveform-level resource scheduling, signal-level joint processing, and scenario adaptive configuration.

\section{Conclusions}

In this paper, we revisit HPR from an ISAC perspective and highlight 5G-based HPR as a compelling alternative to vision-based, WiFi-based, and radar-based solutions.
Then, we further investigate typical application scenarios of 5G-based HPR in indoor environments.
Additionally, we propose MiDiPose, a 5G-enabled multi-feature domains framework for human pose recognition.
By modeling amplitude, phase, and Doppler information as aligned and semantically consistent feature domains, MiDiPose achieves effective cross-domain interaction and robust pose estimation.
The proposed framework demonstrates the potential of 5G sensing in precise and scalable HPR, while alleviating privacy concerns compared to vision-based solutions.
In the future, deploying 5G-based HPR will require broader pose or scene coverage and stronger generalization, tighter multi-BS synchronization, and waveform co-design to balance sensing fidelity with communication efficiency.

\bibliographystyle{IEEEtran} 
 \bibliography{IEEEabrv,bib}
\section*{Biographies}
\vspace{-10mm}
\begin{IEEEbiographynophoto}
{Haojin Li} received the M.S. degree in information and communication engineering from the University of China Academy of Telecommunication Technology, Beijing, China, in 2020. He joined Sony China Research Laboratory in 2020 as a 5G researcher and since then has been responsible for Sony Group Corporation 5G system level simulator development, in June 2022, he was promoted to Deputy Principal Research and Development Researcher. He is currently working toward the Ph.D. degree in Electronics Information with the School of Computer and Communication Engineering, University of Science and Technology Beijing, Beijing, China. He is currently a Deputy Director of Sony and USTB Joint Laboratory. His research interests include integrated sensing and communication and non-terrestrial networks. 
\end{IEEEbiographynophoto}
\vspace{-10mm}
\begin{IEEEbiographynophoto}
{Dongzhe Li} is currently pursuing the M.S. degree in Communication Engineering at the University of Science and Technology Beijing, Beijing, China. His research interests include integrated sensing and communication.
\end{IEEEbiographynophoto}
\vspace{-10mm}
\begin{IEEEbiographynophoto}
{Anbang Zhang} received the M.S. degree with Shandong University, Jinan, China. His research interests include semantic communications and machine learning.
\end{IEEEbiographynophoto}
\vspace{-10mm}
\begin{IEEEbiographynophoto}
{Wenqi Zhang} received the B.S. and Ph.D. degrees from Beijing University of Posts and Telecommunications, Beijing, China, in 2016 and 2021, respectively. From June 2021 to August 2023, she was engaged in post-doctoral research with the School of Cyberspace Security, Beijing University of Posts and Telecommunications. She is currently a researcher with the Sony (China) Research Laboratory. Her research interests include intelligent transportation systems, reinforcement learning, and federated learning.
\end{IEEEbiographynophoto}
\vspace{-10mm}
\begin{IEEEbiographynophoto}
{Chen Sun} received the Ph.D. degree in electrical engineering from Nanyang
Technological University, Singapore, in 2005. From August 2004 to May
2008, he was a researcher at ATR Wave Engineering Laboratories, Japan
working on adaptive beamforming and direction finding algorithms of parasitic
array antennas as well as theoretical analysis of cooperative wireless networks.
In June 2008, he joined the National Institute of Information and Communications
Technology (NICT), Japan, as expert researcher working on distributed
sensing and dynamic spectrum access in TV white space. He is
currently the deputy head of Beijing Lab at Sony R\&D Center. His research
interests include smart antennas, cognitive radio, V2X, FL and AI in wireless
networks.
\end{IEEEbiographynophoto}
\vspace{-10mm}
\begin{IEEEbiographynophoto}
{Haijun Zhang} is currently a Full Professor at University of Science and
Technology Beijing, China. He was a Postdoctoral Research Fellow in
Department of Electrical and Computer Engineering, the University of British
Columbia (UBC), Canada. He serves/served as Track Co-Chair of VTC
Fall 2022 and WCNC 2020/2021, Symposium Chair of Globecom’19, TPC
Co-Chair of INFOCOM 2018 Workshop on Integrating Edge Computing,
Caching, and Offloading in Next Generation Networks, and General Co-Chair
of GameNets’16. He serves/served as an Editor of IEEE Transactions on Communications,
and IEEE Transactions on Network Science and Engineering. He
received the IEEE CSIM Technical Committee Best Journal Paper Award in
2018, IEEE ComSoc Young Author Best Paper Award in 2017, IEEE ComSoc
Asia-Pacific Best Young Researcher Award in 2019. He is a Distinguished
Lecturer of IEEE and a Fellow of IEEE.
\end{IEEEbiographynophoto}

\end{document}